\documentclass[pra,twocolumn,amsmath,amssymb,superscriptaddress,showpacs]{revtex4}
\usepackage[latin9]{inputenc}
\usepackage{amsmath}
\usepackage{amssymb}
\usepackage{epsfig}

\usepackage{ucs}

\usepackage{bbm}

\newcommand{\trace}{{\rm Tr}}

\usepackage{array}
\makeatletter
\newcommand{\thickhline}{%
    \noalign {\ifnum 0=`}\fi \hrule height 1pt
    \futurelet \reserved@a \@xhline
}
\newcolumntype{"}{@{\hskip\tabcolsep\vrule width 1pt\hskip\tabcolsep}}
\makeatother

\begin{document}

\title{Extremal properties of the variance and the quantum Fisher information}

\begin{abstract}
We show that  the variance is its own concave roof. 
For rank-$2$ density matrices and operators with zero 
diagonal elements in the eigenbasis of the density matrix, we prove analytically
that the quantum Fisher information is four times the convex roof of the variance.
Strong numerical evidence suggests that this statement is true
even for operators with nonzero
diagonal elements or density matrices with a rank larger than $2.$
We also find that within the different types of generalized quantum Fisher information
considered in [D. Petz, J. Phys. A: Math. Gen. \textbf{35}, 929 (2002);
P. Gibilisco, F. Hiai, and D. Petz, IEEE Trans. Inf.
Theory \textbf{55}, 439 (2009)], after appropriate normalization,
the quantum Fisher information is the largest.
Hence, we conjecture that the quantum Fisher information is four times the convex roof of the variance
even for the general case.
\end{abstract}

\author{G\'eza T\'oth}
\affiliation{Department of Theoretical Physics, University of the Basque
Country UPV/EHU, P.O. Box 644, E-48080 Bilbao, Spain}
\affiliation{IKERBASQUE, Basque Foundation for Science, E-48011 Bilbao, Spain}
\affiliation{Wigner Research Centre for Physics, Hungarian
Academy of Sciences, P.O. Box 49, H-1525 Budapest, Hungary}
\email{toth@alumni.nd.edu}
\homepage[\\URL: ]{http://www.gtoth.eu}
\author{D\'enes Petz}
\affiliation{Alfr\'ed R\'enyi Institute of Mathematics, Re\'altanoda utca 13-15, H-1051
Budapest, Hungary}
\affiliation{Department of Mathematical Analysis,
Budapest University of Technology and Economics, 
H-1111 Budapest, Hungary}


\maketitle


\date{\today}

\global\long\def\ket#1{\vert#1\rangle}

\global\long\def\ketbra#1{\vert#1\rangle\langle#1\vert}

\global\long\def\exs#1{\langle#1\rangle}

\global\long\def\WW{\mathcal{W}}

\global\long\def\va#1{{\ensuremath{(\Delta#1)^{2}}}}

\global\long\def\varrm{{\ensuremath{{\rm var}}}}

\global\long\def\varrmh{{\ensuremath{{\rm \hat{var}}}}}

\global\long\def\openone{\mathbbm{1}}

\section{Introduction}
Metrology plays an important role in many areas of physics and engineering. 
Quantum metrology \cite{GL04,GL06} is a novel development, which takes into account the quantum
nature of matter when determining the achievable maximal accuracy in parameter estimation.
For instance, quantum metrology shows that it is possible to reach a much higher accuracy in typical metrological tasks in a linear interferometer with highly entangled quantum systems  than with non-entangled ones, in which no real quantum interaction took place between the particles
 \cite{GT09,SD01,SM01,PS01,
 FisherMultipartieEntanglement,FisherMultipartieEntanglement2,FisherMultipartieEntanglement3}.
 In practical situations, the influence of noise can 
 reduce considerably the
 gain in accuracy when applying entangled quantum states for metrology \cite{DK12}.

\begin{figure}
\centerline{\epsfxsize=3.5in\epsffile{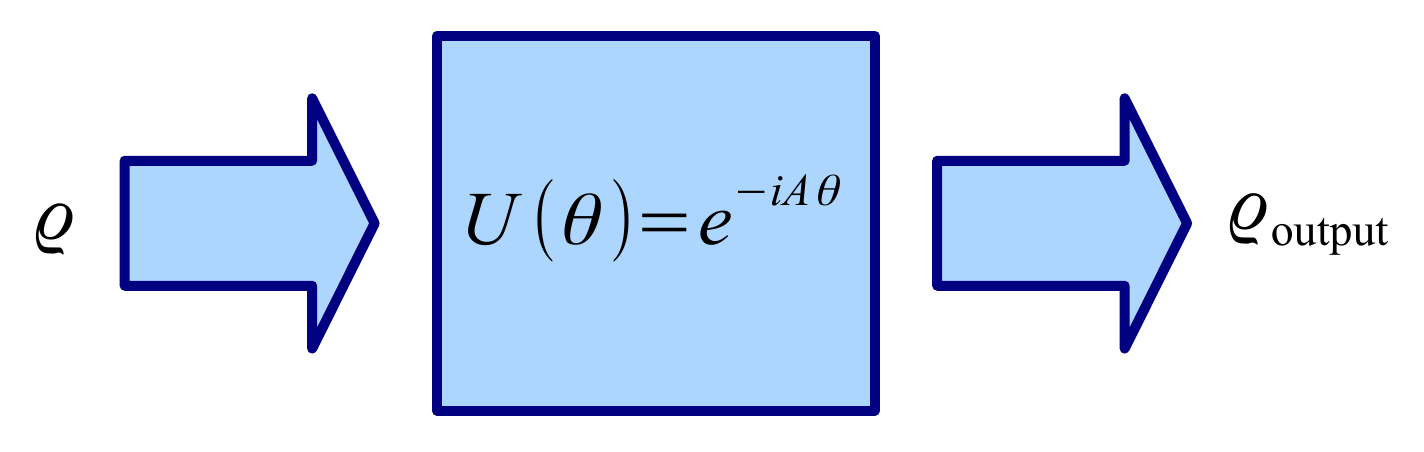}
} \caption{(Color online) 
 Basic parameter estimation task in quantum metrology. The small parameter $\theta$ must be estimated by making
 measurements on the output state $\varrho_{\rm output}.$}
\label{fig_interfero}
\end{figure}

Quantum variance and quantum Fisher information are two fundamental notions of quantum metrology.
When measuring a quantum mechanical observable $A,$ the uncertainty of the measurement outcome is described by the variance defined as
\begin{equation}
\va A_{\varrho}=\exs{A^{2}}_{\varrho}-\exs A_{\varrho}^{2},\label{eq:usualvar}
\end{equation}
where $A$ is a Hermitian operator and $\varrho$ is a density matrix.
Connected to this uncertainty, when we would like to estimate a parameter of a quantum evolution based on measuring quantum observables, an uncertainty arises also in this parameter estimation.
The most fundamental parameter estimation task is estimating the small parameter $\theta$ of the unitary dynamics $U=\exp(-iA\theta)$
via making measurements on the output state (see Fig.~\ref{fig_interfero}). 
The accuracy of estimating $\theta$ is bounded from below by the famous Cram\'er-Rao bound as
\begin{equation}
\Delta \theta \ge \frac{1}{\sqrt{F_{Q}^{{\rm BC}}[\varrho,A] }},
\end{equation}
where 
the quantum Fisher information is defined as \cite{PS01,FisherMultipartieEntanglement,FisherMultipartieEntanglement2,FisherMultipartieEntanglement3,BC94,H82,H76,P08}
\begin{equation}
F_{Q}^{{\rm BC}}[\varrho,A]=2\sum_{i,j}\frac{(\lambda_{i}-\lambda_{j})^{2}}{\lambda_{i}+\lambda_{j}}\vert A_{ij}\vert^{2}.\label{eq:qF}
\end{equation}
Here $\lambda_i$ are the eigenvalues of the density matrix and $A_{ij}$ are the matrix elements of the operator $A$ in the eigenbasis of the density matrix.

In Refs.~\cite{P03,GH09,DG11}, it has been shown that it is possible to define
generalized variances and various types of generalized quantum Fisher information.
In fact, the two notions are closely connected to each other as for every generalized
variance there is a corresponding generalized quantum Fisher information.
Thus, the question arises: Are the usual variance, Eq.~(\ref{eq:usualvar}), and 
  the usual quantum Fisher information, Eq.~(\ref{eq:qF}), 
special among the generalized quantities? In this paper we will answer these questions
affirmatively. Our findings show that fundamental quantities used in metrology
have extremal properties. 

In particular, in the first part of the paper we will define generalized variances.
The definition presented is broader than that of  Refs.~\cite{P03,GH09,DG11}. Then,
we will show that the usual variance is the smallest generalized variance and, connected to this fact,
it fulfills the following important relation.

\textbf{Theorem 1.}\emph{ Let us denote the decomposition of a  density matrix into
the mixture of pure states as
\begin{equation}
\varrho=\sum_k p_k \ketbra{\Psi_k}. \label{decomp}
\end{equation}
Then, the variance for a mixed state can be obtained through  a concave roof construction as
\begin{equation}
\va A_{\varrho}=\sup_{\{ p_k,\ket{\Psi_k}\}} \sum_k p_k \va A_{\Psi_k}. 
\end{equation}
The proof of Theorem 1 will be given later.}

In the second part
of the paper we define the generalized quantum Fisher information.
The definition presented is also broader than that of  Refs.~\cite{P03,GH09,DG11}.
We will show that $F_{Q}^{\rm BC}$ defined in Eq.~\eqref{eq:qF} is the largest 
among the types of generalized quantum Fisher information
for the low rank case, and connected to this fact, it fulfills the following relation.

\textbf{Theorem 2.} \emph{For rank-$2$ 
density matrices and for an $A$ with zero diagonal elements in the eigenbasis of the density matrix, 
the quantum Fisher information can be given as the convex roof of the variance as
\begin{equation}
F_{Q}^{\rm BC}[\varrho,A] =4 \inf_{\{ p_k,\ket{\Psi_k}\}} \sum_k p_k \va {A
 }_{\Psi_k},\label{e2b}
\end{equation}
where $\{p_{k},\ket{\Psi_{k}}\}$ refers to a decomposition of $\varrho$ of the type Eq.~(\ref{decomp}).
The proof of Theorem 2 will also be given later.}

We will also discuss that numerical calculations suggest that
the left-hand side and the right-hand side of Eq.~(\ref{e2b})
are very close to each other, even
when $A$ has non-zero diagonal elements or the density matrix has a rank larger than
$2.$ 

Concerning the quantum Fisher information, we can also prove the following theorem.

\textbf{Theorem 3.}  {\it $F_{Q}^{\rm BC}[\varrho,A]$  defined in Eq.~\eqref{eq:qF} is maximal within the
various types of quantum Fisher information defined by Refs.~\cite{P03,GH09,DG11}, 
if they are normalized such that for pure states they equal $4(\Delta A)^2.$}

Concerning Theorem 3, it is important to note that all types of quantum 
Fisher information defined by Refs.~\cite{P03,GH09,DG11} are convex.
Based on Theorem 3 and the strong numerical evidence mentioned above,
 we formulate the following conjecture.

\textbf{Conjecture 1.} \emph{The quantum Fisher information $F_{Q}^{\rm BC}[\varrho,A] $ 
defined in Eq.~\eqref{eq:qF} is 
4 times the convex roof of the variance for density matrices of any rank and any observable $A$.}

The statements of Theorem 1 and Conjecture 1 can be concisely reformulated as follows.
For any decomposition $\{p_{k},\ket{\Psi_{k}}\}$ of the density matrix $\varrho$ we have
\begin{equation}
\frac{1}{4}F_{Q}^{\rm BC}[\varrho,A]  \le  \sum_k p_k \va {A}_{\Psi_k} \le \va A_{\varrho},
\end{equation}
where the upper and the lower bounds are both tight.

The paper is organized as follows. In
Sec.~II, we define generalized variances and determine the minimal generalized variance. In Sec.~III,
we consider the generalized quantum Fisher information for the rank-$2$ case, and look for the maximal
generalized quantum Fisher information.
In Sec.~IV, we present numerical calculations for density matrices with a higher rank.
In Sec.~V, we show that $F_{Q}^{\rm BC}$ is the largest among the types of quantum Fisher information defined by Refs. \cite{P03,GH09,DG11}. 
In Appendix A, we give some details of the semidefinite programs needed for Sec.~IV.

\section{Generalized variances and the proof of Theorem 1}

In this section, we will define the generalized variance and the minimal
generalized variance. We will also show that the minimal generalized
variance is the usual variance given in Eq.~\eqref{eq:usualvar}.

\textbf{Definition 1.} {The generalized variance ${\rm var}_{\varrho}(A)$
is defined by the following two requirements.
\begin{enumerate}
\item For pure states, the generalized variance equals the usual variance 
\begin{equation}
{\rm var}{}_{\Psi}(A)=\va A_{\Psi}.
\end{equation}
\item For mixed states, $\varrm_{\varrho}(A)$ is concave in the state.
\end{enumerate}

There are infinitely many generalized variances that fulfill the requirements
of Definition 1. Next, we will show that it is possible to identify
the smallest one. }

\textbf{Definition 2.} {The minimal generalized variance
$\varrm{}_{\varrho}^{\min}(A)$ is defined by  the following  two requirements.
\begin{enumerate}
\item For pure states, it equals the usual variance
\begin{equation}
{\rm var}{}_{\Psi}^{\min}(A)=\va A_{\Psi},
\end{equation}
\item For mixed states, it is defined through a \emph{concave roof} construction
\cite{roofs}
\begin{equation}
{\rm var}{}_{\varrho}^{\min}(A)=\sup_{\{p_{k},\ket{\Psi_{k}}\}}\sum_{k}p_{k}\va A_{\Psi_{k}},
\label{sup}
\end{equation}
where $\{p_{k},\ket{\Psi_{k}}\}$ refers to a decomposition of $\varrho$ of the type Eq.~(\ref{decomp}).
\end{enumerate}}

Note the counterintuitive fact that while a supremum is used to define Eq.~\eqref{sup}, it is the minimal generalized variance and not the maximal one. Let us show that it is indeed the minimal generalized variance.
It is clear that for any generalized variance, we have
\begin{equation}
\varrm{}_{\varrho}(A)\ge\varrm{}_{\varrho}^{\min}(A).\label{mins}
\end{equation}
If this were not true then it would be possible that for some 
decomposition of the type Eq.~\eqref{decomp}
we have ${\rm var}{}_{\varrho}(A)<\sum_{k}p_{k} {\rm var}{}_{\Psi_{k}}(A),$
which would contradict the concavity of the variance.
On the other hand, Eq.~\eqref{sup} is concave in the state, 
since it is defined by a concave roof construction, thus
it fulfills Definition 1.
Hence, Eq.~\eqref{sup} is the minimal generalized variance.

Next we will prove two lemmas that we need later in the proof of the central
theorem of the section.

\textbf{Lemma 1.} \emph{For rank-$2$ states, the minimal
generalized variance is the usual variance given in Eq.~\eqref{eq:usualvar}.
In particular, this statement is true for all qubit states.}

\emph{Proof.} For all decompositions of the form Eq.~\eqref{decomp}, the minimal variance is bounded from above and below as
\begin{equation}
\va A_{\varrho}\ge\varrm{}_{\varrho}^{\min}(A) \ge \sum_{k}p_{k}\va A_{\Psi_{k}}.\label{gee}
\end{equation}
The first inequality in Eq.~\eqref{gee} is due to Eq.~\eqref{mins}, the second one is due to the concavity of the variance. 

In order to proceed to prove Lemma 1, we need to know 
an important property of the usual variance $\va A.$
It can be written in the following way
\begin{equation} \label{deccc}
\va A_{\varrho}=\sum_{k}p_{k}\left[\va A_{\Psi_{k}}+(\exs A_{\Psi_{k}}-\exs A_{\varrho})^{2}\right].
\end{equation}
Here the summation is over the pure subensembles.
$\va A_{\Psi_{k}}$ is the variance in the $k$th subensemble, 
 while the $(\exs A_{\Psi_{k}}-\exs A_{\varrho})^{2}$ term
is the square of the difference between the expectation value for the subensemble 
and  the expectation value for the entire ensemble. One can call the two terms the "quantum" 
and the "classical" part of the variance, respectively. 

Based on Eq.~\eqref{deccc} we can state the following.
If for a quantum state $\tilde{\varrho}$ there is a decomposition $\{\tilde{p}_k,\ket{\tilde{\Psi}_k}\}$
such that the subensemble expectation values equal the expectation value for the entire ensemble 
(i.e., $\exs A_{\tilde{\Psi}_{k}}=\exs A_{\tilde{\varrho}}$ for all $k$)
then for such a decomposition we have
\begin{equation}
\va A_{\tilde{\varrho}} = \sum_{k}\tilde{p}_{k}\va A_{\tilde{\Psi}_{k}}.\label{gee2}
\end{equation}
Based on Eq.~\eqref{gee}, we can now state that
for $\tilde{\varrho}$ the minimum generalized variance of $A$ and the 
usual variance are equal
\begin{equation}
\va A_{\tilde{\varrho}}=\varrm{}_{\tilde{\varrho}}^{\min}(A). \label{geerrr}
\end{equation}

We will now show that such a decomposition, in which the subensemble 
expectation values are all equal, always exists
for rank-2 density matrices. Hence, we will show that for all such states
the usual quantum variance is the minimal generalized variance.

\begin{figure}
\centerline{\epsfxsize=2.7in\epsffile{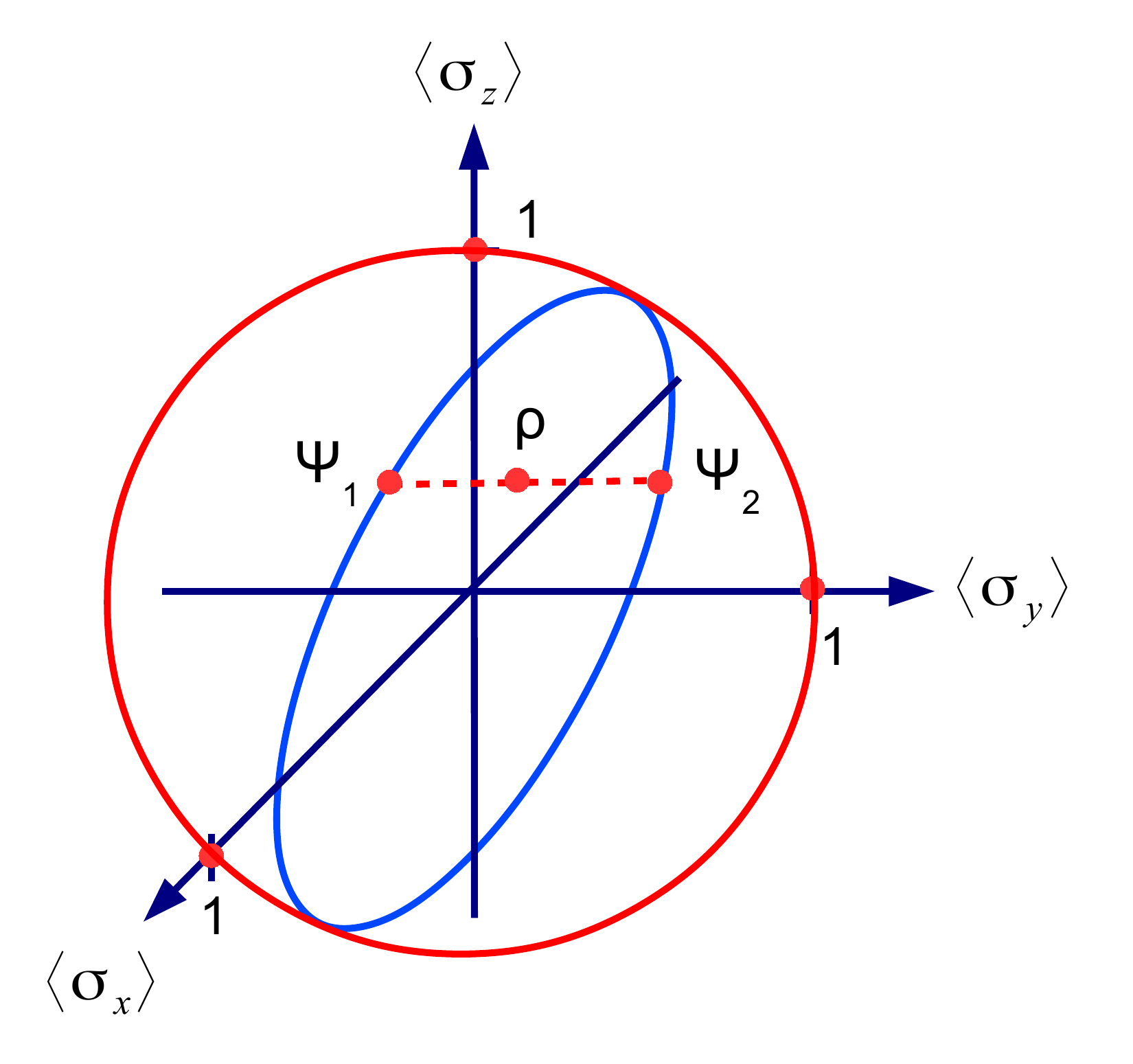}
} 
\caption{ (Color online) 
A single-qubit mixed state $\varrho$ is decomposed as 
Eq.~\eqref{xyzt} into the mixture of pure states $\ket{\Psi_{\phi_{1}}}$ and
$\ket{\Psi_{\phi_{1}+\pi}}$ defined in the text. Quantum states are represented by points in the
$(\exs{\sigma_x},\exs{\sigma_y},\exs{\sigma_z})-$space.
The sphere with radius $1$ is the Bloch sphere.
Density matrices $\varrho'$ for which the expectation value of $A$ is the same 
as for $\varrho$ fulfill the linear condition
${\rm Tr}[A(\varrho-\varrho')]=0$ and are on a plane. The elliptic curve is the cross section between this plane and the surface of the Bloch sphere. Points corresponding to $\varrho,$ $\ket{\Psi_{\phi_{1}}}$ and $\ket{\Psi_{\phi_{1}+\pi}}$ are all on this plane, thus the operator $A$ has the same expectation value in these three states. 
}
\label{fig_rank2}
\end{figure}

Let us take an eigendecomposition of the state $\varrho$
\begin{equation}
\varrho=p\ketbra{\Psi_{1}}+(1-p)\ketbra{\Psi_{2}}.
\end{equation}
Thus, in the $\ket{\Psi_{1}}$ and $\ket{\Psi_{2}}$ basis we can
write the density matrix $\varrho$  as
\begin{equation}
\varrho=\left[\begin{array}{cc}
p & 0\\
0 & 1-p
\end{array}\right].
\end{equation}
Let us define now the family of states
\begin{equation}
\ket{\Psi_{\phi}}=\sqrt{p}\ket{\Psi_{1}}+\sqrt{1-p}\ket{\Psi_{2}}e^{i\phi}. \label{PPP}
\end{equation}
For the states (\ref{PPP}), the expectation value of the operator $A$ can be written as
\begin{align}
\exs{\Psi_{\phi}\vert A\vert\Psi_{\phi}}
= & \exs A_{\varrho}+2\sqrt{p(1-p)}{\rm Re}\left(\exs{\Psi_{1}\vert A\vert\Psi_{2}}e^{i\phi}\right).\label{ReRe}
\end{align}
Clearly, there is an angle $\phi_{1}$ such that 
\begin{equation}
{\rm Re}\left(\exs{\Psi_{1}\vert A\vert\Psi_{2}}e^{i\phi_{1}}\right)=0.
\end{equation}
For this angle, the expectation value  of $A$ in the state $\ket{\Psi_{\phi_1}}$ 
is the same as its expectation value in $\varrho,$
since the second term in Eq.~(\ref{ReRe}) is zero.
In the  basis of the states $\ket{\Psi_{1}}$ and $\ket{\Psi_{2}},$ we can write the projection operators onto $\ket{\Psi_{\phi_1}}$ as
\begin{align}
\ketbra{\Psi_{\phi_{1}}}\nonumber \\
= & \left[\begin{array}{cc}
p & \sqrt{p(1-p)}e^{-i\phi_{1}}\\
\sqrt{p(1-p)}e^{+i\phi_{1}} & 1-p
\end{array}\right]. \label{e1}
\end{align}
The same is true for $\phi_{1}+\pi.$ In the  basis of the states $\ket{\Psi_{1}}$ and $\ket{\Psi_{2}},$
we can write  the projection operator onto $\ket{\Psi_{\phi_1+\pi}}$ as
\begin{align}
\ketbra{\Psi_{\phi_{1}+\pi}}\nonumber\\
= & \left[\begin{array}{cc}
p & -\sqrt{p(1-p)}e^{-i\phi_{1}}\\
-\sqrt{p(1-p)}e^{+i\phi_{1}} & 1-p
\end{array}\right]. \label{e2}
\end{align}
Based on Eqs.~(\ref{e1}) and (\ref{e2}), the state $\varrho$ can be decomposed as 
\begin{equation} \label{xyzt}
\varrho=\frac{1}{2}\left(\ketbra{\Psi_{\phi_{1}}}+\ketbra{\Psi_{\phi_{1}+\pi}}\right),
\end{equation}
and the two component states have $\exs{\Psi_{\phi_{1}}\vert A\vert\Psi_{\phi_{1}}}=\exs A_{\varrho}$
and $\exs{\Psi_{\phi_{1}+\pi}\vert A\vert\Psi_{\phi_{1}+\pi}}=\exs A_{\varrho},$ as expected.
\ensuremath{\hfill \Box}

In Fig.~\ref{fig_rank2}, we show an example to demonstrate how Lemma 1 works for single-qubit states. The three coordinate axes are the expectation values of the three Pauli spin matrices. Points on the surface of the Bloch ball correspond to pure states, points within the ball correspond to rank-$2$ mixed states. The operator  $A$ has the same
expectation value in the states $\varrho,$ $\ket{\Psi_{\phi_{1}}}$ and $\ket{\Psi_{\phi_{1}+\pi}}.$ Note that ${\sigma_z}$ has also the same expectation value in these states.

Next we will consider the decomposition of states with a rank larger than $2.$

\textbf{Lemma 2.} \emph{Let us consider an eigendecomposition
of a density matrix 
\begin{equation} \label{declambda}
\varrho_{0}=\sum_{k=1}^{r_{0}}\lambda_{k}\ketbra{\Psi_{k}}
\end{equation}
 with all $\lambda_{k}>0.$ Let us denote the rank of the density matrix as $r(\varrho_{0})=r_{0}.$
We consider density matrices for which $r_{0}\ge3.$ Let us denote by $A_{0}$
the expectation value of an operator $A$ in the state $\varrho_{0}$ 
\begin{equation}
{\rm Tr}(A\varrho_{0})=A_{0.}\label{eq:trace}
\end{equation}
 We claim that for any $A,$ $\varrho_{0}$ can always be decomposed
as
\begin{equation}
\varrho_{0}=p\varrho_{-}+(1-p)\varrho_{+},\label{eq:decomp}
\end{equation}
 such that $r(\varrho_{-})<r_{0},$ $r(\varrho_{+})<r_{0},$ and
\begin{equation}
{\rm Tr}(A\varrho_{+})={\rm Tr}(A\varrho_{-})=A_{0}.
\end{equation}}

\emph{Proof.} Let us consider the eigenvalues $\lambda_k$
in the eigendecomposition of $\varrho,$ Eq.~\eqref{declambda}.
First, note that from the conditions ${\rm Tr}(\varrho)=1$ and Eq.~\eqref{eq:trace}, it follows for the $\lambda_k$'s that
\begin{subequations}
\begin{eqnarray}
\sum_{k=1}^{r_0}\lambda_{k}&=&1,\label{summalambda}\\
\sum_{k=1}^{r_0}\lambda_{k}a_k&=&A_{0},\label{eq:trace2}
\end{eqnarray}
\end{subequations}
where we introduced for the subensemble expectation values the notation
\begin{equation}
a_{k}=\langle\Psi_{k}\vert A\vert\Psi_{k}\rangle.\label{eq:trace3}
\end{equation}

Let us now define a family of states as
\begin{equation}
\varrho(c,\Delta \vec \lambda)=\sum_{k=1}^{r_{0}}(\lambda_{k}+c\Delta\lambda_{k})\ketbra{\Psi_{k}},\label{eq:rhoc}
\end{equation}
where $c$ is a real parameter and  $\Delta\lambda_{k}$ fulfill
\begin{subequations}
\begin{eqnarray}
\sum_{k=1}^{r_0}\Delta\lambda_{k}&=&0,\label{eq:a2}\\
\sum_{k=1}^{r_0}a_{k}\Delta\lambda_{k}&=&0.\label{eq:a1}
\end{eqnarray}
\end{subequations}
Equation~\eqref{eq:a2} ensures that the state $\varrho(c,\Delta \vec \lambda)$ has a unit trace, while 
due to Eq.~\eqref{eq:a1} for all $c$
\begin{equation}
{\rm Tr}[A\varrho(c,\Delta \vec \lambda)]=A_{0.}
\end{equation} 
We can rewrite Eqs.~\eqref{eq:a2} and \eqref{eq:a1} as
\begin{equation}\label{twolines}
\left[\begin{array}{ccccc}
a_{1} & a_{2} & ... & a_{r_{0}-1} & a_{r_{0}}\\
1 & 1 & ... & 1 & 1
\end{array}\right]\Delta\vec{\lambda}=\left[\begin{array}{c}
0\\
0
\end{array}\right].
\end{equation}
For $r_{0}\ge3,$ there is always a $\Delta\vec{\lambda}\ne0$ fulfilling
Eq.~\eqref{twolines}. It is clear that, if
$\varrho(c,\Delta \vec \lambda)$ has non-negative eigenvalues, then $\varrho(c,\Delta \vec \lambda)$ is
a physical state. In the following, for compactness, we will omit $\Delta \vec \lambda$ from the argument 
of $\varrho.$

Let us now examine the properties of the $\varrho(c)$ family. It
is clear that
\begin{equation}
\varrho(0)=\varrho_{0}.
\end{equation}
Moreover, it is also clear that around $c=0$ there is some range
of values for $c,$ such that $\varrho(c)$ is physical. So first,
let us start to increase $c$ gradually from zero until one of the
eigenvalues becomes zero. Let us denote this value by $c_{+}.$ Thus,
$\varrho(c)$ is physical for 
\begin{equation}
0\le c\le c_{+}.
\end{equation}
We have $r(\varrho(c_{+}))<r_{0}.$ Moreover, $\varrho(c)$ is non-physical
for $c>c_{+}.$ A similar thing happens, if we start from $c=0$ by
decreasing $c$ gradually until one of the eigenvalues becomes zero.
Let us denote this value by $c_{-}.$ Thus, $\varrho(c)$ is physical
for 
\begin{equation}
c_{-}\le c\le0.
\end{equation}
Again, we have $r(\varrho(c_{-}))<r_{0}.$ Moreover, $\varrho(c)$
is non-physical for $c<c_{-}.$ In summary, for any choice of $\varrho$
and $\Delta\lambda_{k}$ fulfilling Eqs.~\eqref{eq:a1} and \eqref{eq:a2},
there is a family of states $\varrho(c)$ defined in Eq.~\eqref{eq:rhoc}
such that $\varrho(c)$ is physical for 
\begin{equation}
c_{-}\le c\le c_{+},
\end{equation}
and $r(\varrho(c_{+}))<r_{0},$ and $r(\varrho(c_{-}))<r_{0}.$ One
can explicitly reconstruct these bounds as
\begin{equation}
c_{+}=\min_{k}\frac{\lambda_{k}}{\Theta(-\Delta\lambda_{k})}
\end{equation}
and
\begin{equation}
c_{-}=-\min_{k}\frac{\lambda_{k}}{\Theta(\Delta\lambda_{k})},
\end{equation}
where 
\begin{equation}
\Theta(x)=\begin{cases}
x & \text{\text{if }}x\ge0,\\
0 & \text{if }x<0.
\end{cases}
\end{equation}

Let us now construct the decomposition Eq.~\eqref{eq:decomp}. Direct
calculations can show that the following decomposition is of the desired form
\begin{equation}
\varrho_{-}=\varrho(c_{-}),
\end{equation}
and
 \begin{equation}
\varrho_{+}=\varrho(c_{+}),
\end{equation}
while the mixing probability is
\begin{equation}
p=\frac{c_{+}}{c_{+}-c_{-}}.
\end{equation}
\ensuremath{\hfill \Box}

\begin{figure}
\centerline{\epsfxsize=2.9in\epsffile{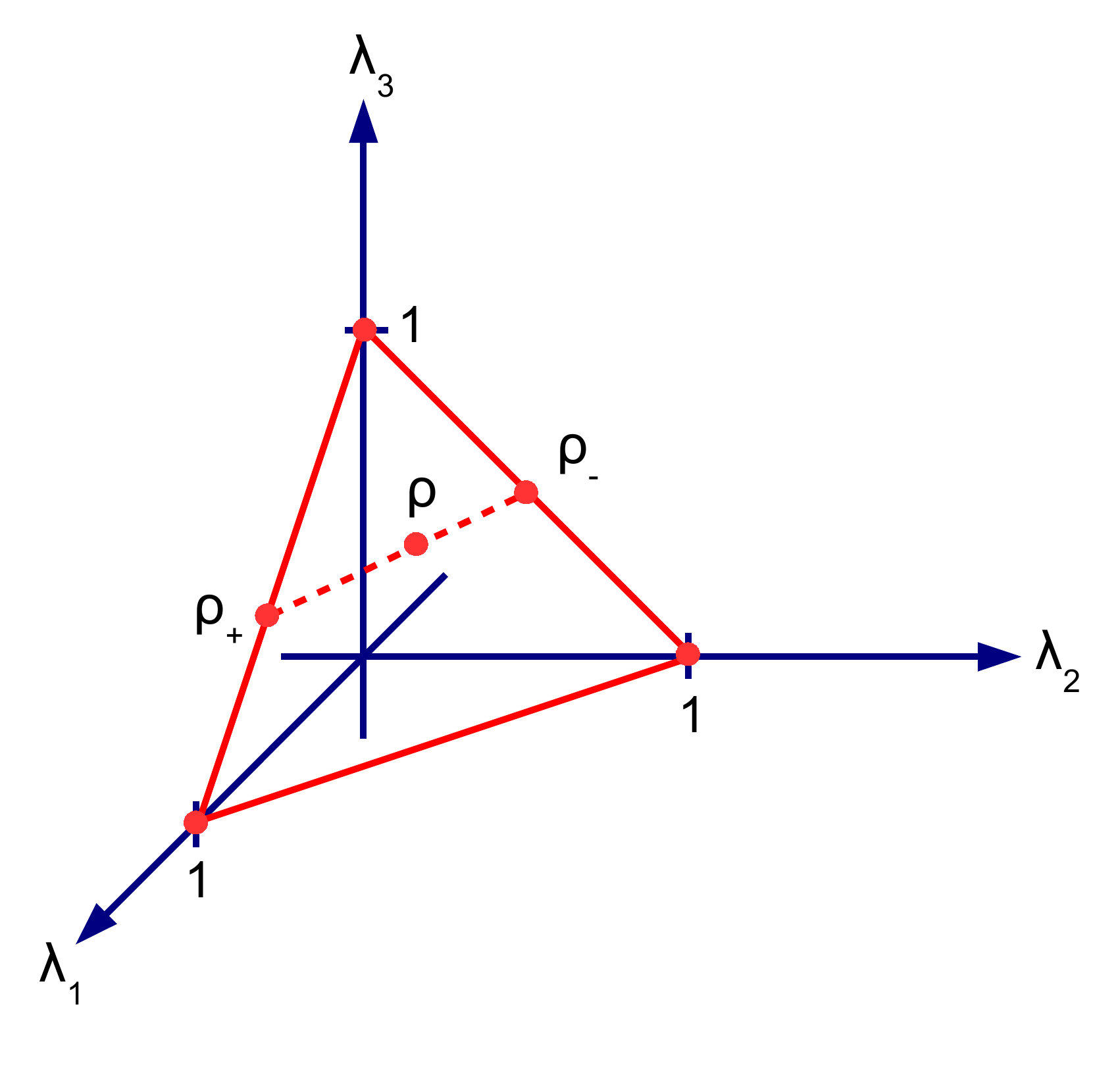}
} \caption{(Color online) The rank-3 mixed state $\varrho_0$ is decomposed as Eq.~(\ref{eq:decomp}) into the mixture of two rank-2 states, $\varrho_-$ and $\varrho_+.$ The coordinate axes are the eigenvalues of the density matrix, while we assume that all density matrices have the same eigenstates as $\varrho.$
Points corresponding to states $\varrho'$ for which ${\rm Tr}(\varrho'A)={\rm Tr}(\varrho_0 A)$ are on the dashed line.
For $\varrho_+$ we have $\lambda_2=0, $ while for $\varrho_-$ we have $\lambda_1=0.$
}
\label{fig_rank3}
\end{figure}

In Fig.~\ref{fig_rank3}, we presented an example showing how Lemma 2  works for a rank-$3$ state.
A point $(\lambda_1,\lambda_2,\lambda_3)$ 
in the coordinate system corresponds to $\varrho=\sum_{k=1}^3\lambda_k\ketbra{\Psi_k}.$
All points have to fulfill Eq.~(\ref{summalambda}) and $\lambda_k\ge 0$ for $k=1,2,3.$ Thus, all points corresponding to physical states are in the triangle. Interior points of the triangle correspond to rank-$3$ states, while points on the boundary correspond to rank-$2$ and rank-$1$ states.

After proving Lemma 1 and Lemma 2, we will prove the main theorem
of the section.

\emph{Proof of Theorem 1.} Theorem 1 is equivalent to the statement that 
for systems of any dimension, the
minimal generalized variance is the usual variance 
\begin{equation}
\varrm{}_{\varrho}^{\min}(A)=\va A_{\varrho}.
\end{equation}
For rank-$2$ states, this is true because of Lemma 1.
According to Lemma 2, any state with a rank larger than 2 can be decomposed
into the mixture of lower rank states that have the same expectation
value for $A$ as the original state has. The lower rank states can
then be decomposed into the mixture of states with an even lower rank,
until we reach rank-$2$ states. Thus, any state $\varrho$ can be
decomposed into the mixture of the form Eq.~\eqref{decomp}
such that $\langle\Psi_{k}\vert A\vert\Psi_{k}\rangle={\rm Tr}(A\varrho).$
Hence, similarly ro the proof of Lemma 1, the statement of Theorem
1 follows \cite{alternative}. \ensuremath{\hfill \Box}

We will present two examples in order to clarify certain issues concerning
generalized variances.

\textbf{Example 1. }Let us consider the state 
\begin{equation}
\varrho=\frac{1}{2}\left(\ketbra {+1}+\ketbra{-1}\right),
\end{equation}
where $\ket{\pm 1}$ denote the eigenstates of the Pauli spin matrix $\sigma_z.$
For this state, we have $\va{\sigma_{z}}=+1.$ Let us look at the
decomposition 
\begin{enumerate}
\item $p_{1}=\frac{1}{2},\vert\Psi_{1}\rangle=\frac{1}{\sqrt{2}}(\ket {+1}+\ket {-1}),$
\item $p_{2}=\frac{1}{2},\vert\Psi_{2}\rangle=\frac{1}{\sqrt{2}}(\ket {+1}-\ket {-1}).$ 
\end{enumerate}
For this decomposition $\va{\sigma_{z}}_{\Psi_{k}}=+1$ and $\langle\Psi_{k}\vert\sigma_{z}\vert\Psi_{k}\rangle={\rm Tr}(\sigma_{z}\varrho)=0.$ Thus, we presented a decomposition such that the expectation value of $A$ is the same for both subensembles $\ket{\Psi_{k}}.$

\textbf{Example 2. }Let us consider the generalized variance
\begin{equation}
\varrm^{{\rm quadratic}}_{\varrho}(A)={\rm Tr}(A^{2}\varrho)-{\rm Tr}(A\varrho)^{2}+1-{\rm Tr}(\varrho^2).
\label{eq:example2}
\end{equation}
For pure states, it equals the usual variance. For mixed states, it
is larger than that and it is concave in the state. It is in fact
a sum of the usual variance and the linear entropy, which is
zero for pure states and is concave in the state. 
 
\section{Generalized quantum Fisher information and the proof of Theorem 2}

In this section, we will define the generalized quantum Fisher information
and the maximal generalized quantum Fisher information. We will show
that for rank-$2$ density matrices, the maximal quantum Fisher information
is the usual quantum Fisher information.

\textbf{Definition 3.} The generalized quantum Fisher information
$F_{Q}[\varrho,A]$ is defined by  the following  two requirements.
\begin{enumerate}
\item For pure states, we have 
\begin{equation}
F_{Q}[\ketbra{\Psi},A]=4\va {A_{\Psi}}. 
\end{equation}
The factor $4$ appears to keep the consistency with the existing
literature \cite{PS01}.
\item For mixed states, $F_{Q}[\varrho,A]$ is convex in the state.
\end{enumerate}

The usual quantum Fisher information, Eq.~\eqref{eq:qF}, is one of the types of generalized quantum Fisher information.
Next, we will present another well-known generalized quantum Fisher information.

{\bf Example 3.} Let us consider the Wigner-Yanase skew information defined as \cite{WY63}
\begin{equation}
I[\varrho,A]={\rm Tr}(A^2\varrho)-{\rm Tr}(A\varrho^{\frac{1}{2}} A\varrho^{\frac{1}{2}}).
\end{equation}
For pure states it equals $(\Delta A)_\varrho^2$ and it is convex in the state.
Thus  $4I[\varrho,A]$ fulfills Definition 3.

There are infinitely many types of generalized quantum Fisher information that fulfill
the requirements of Definition 3. Next, we will show that it is possible
to identify the largest one.

\textbf{Definition 4.} The maximal generalized quantum Fisher information, 
$F_{Q}^{\max}[\varrho,A]$ is defined
by  the following two requirements.
\begin{enumerate}
\item For pure states, it equals four times the usual variance
\begin{equation}
F_{Q}^{\max}[\ketbra{\Psi},A]=4\va {A}_{\Psi}.
\end{equation}

\item For mixed states, it is defined through a \emph{convex roof} construction
\cite{roofs}
\begin{equation}
F_{Q}^{\max}[\varrho,A]=4\inf_{\{p_{k},\ket{\Psi_{k}}\}} \sum_k p_{k}\va {A}_{\Psi_{k}}.
\label{MAX}
\end{equation}

\end{enumerate}
It is clear that for any generalized quantum Fisher information, we
have
\begin{equation}
F_{Q}[\varrho,A]\le F_{Q}^{\max}[\varrho,A].
\end{equation}
This can be proven similarly as it has been shown that the minimal
generalized variance is smaller than or equal to all other generalized variances.

Next we will present the main theorem of this section.

\emph{Proof of Theorem 2.} Let us consider a density matrix $\varrho$ of rank $2$
given as
\begin{equation}
\varrho=q\ketbra{\phi_{1}}+(1-q)\ketbra{\phi_{2}}, \label{rho}
\end{equation}
and a self-adjoint observable $A.$ We will show that for any $\varrho$
and $A$ exists a decomposition into a mixture of projectors
of the form Eq.~\eqref{decomp}
such that
\begin{equation}
F_{Q}^{{\rm BC}}[\varrho,A]=4\sum_{k}p_{k}\va {A}_{\Psi_{k}}.\label{eq:qqq}
\end{equation}
Knowing that 
\begin{equation}
\sum_{k}p_{k}4\va {A}_{\Psi_{k}}\ge F_{Q}^{{\rm max}}[\varrho,A]\ge F_{Q}^{{\rm BC}}[\varrho,A],
\end{equation}
if a decomposition of the type Eq.~\eqref{eq:qqq}
exists then 
\begin{equation}
F_{Q}^{{\rm BC}}[\varrho,A]=F_{Q}^{\max}[\varrho,A].
\end{equation}
Note that the first inequality in  Eq.~\eqref{eq:qqq} is due to the convexity of the quantum
Fisher information, the second one is due to Definition 4.

We consider the case where the diagonal elements of $A$ are zero, hence $A$
can explicitly be obtained as
\begin{equation}
A= \vert A_{12} \vert \left(\ket{\phi_{1}}\langle\phi_{2}\vert e^{+i\alpha}+
\ket{\phi_{2}}\langle\phi_{1}\vert e^{-i\alpha}\right)
\end{equation}
and $\vert A_{12} \vert$ and $\alpha$ are real.
In fact, the the phase $\alpha$ is irrelevant.
Thus, in our computation only $\vert A_{12}\vert$ is essential, so
we can reduce the problem to the two-dimensional space generated by
$\vert\phi_{1}\rangle$ and $\vert\phi_{2}\rangle,$ and instead of
$A,$ it is sufficient to take into account $\vert A_{12} \vert.$ For
the usual quantum Fisher information, Eq.~\eqref{eq:qF}, we have 
\begin{equation}
F_{Q}^{{\rm BC}}[\varrho,A]=4(2q-1)^{2}\vert A_{12}\vert^{2}.\label{eq:usual}
\end{equation}
For the right-hand side of Eq.~\eqref{eq:qqq} we have
\begin{equation}
4\sum_{k}p_{k}\va{A}_{\Psi_{k}}=4[{\rm Tr}(\varrho A^{2})-\sum_{k}p_{k}
\vert\langle \Psi_k \vert A \vert \Psi_k \rangle\vert^2].\label{eq:sum}
\end{equation}
Based on Eqs.~\eqref{eq:usual} and \eqref{eq:sum},
Eq.~\eqref{eq:qqq} can be rewritten as 
\begin{equation}
4(2q-1)^{2}\vert A_{12}\vert^{2}=4[{\rm Tr}(\varrho A^{2})-
\sum_{k}p_{k} \vert\langle \Psi_k \vert A \vert \Psi_k \rangle\vert^2].\label{eq:www}
\end{equation}
This is the equality to be solved. Using that
\begin{equation}
A^{2}=\vert A_{12}\vert^{2}\left(\ket{\phi_{1}}\langle\phi_{1}\vert+\ket{\phi_{2}}\langle\phi_{2}\vert\right),
\end{equation}
we obtain ${\rm Tr}(\varrho A^{2})=\vert A_{12}\vert^{2}.$
Substituting this into Eq.~\eqref{eq:www}, we get
\begin{equation}
4 q(1-q)\vert A_{12}\vert^{2}=\sum_{k}p_{k}\vert\langle \Psi_k \vert A \vert \Psi_k \rangle\vert^2.\label{eq:qwe}
\end{equation}

Finally, we can present a desired decomposition
 of the form Eq.~\eqref{decomp} for $\varrho$
fulfilling Eq.~\eqref{eq:qwe}. Its is given by two subensembles as 
\begin{equation}
p_{1}=\frac{1}{2},\,\,\,\, \ket{\Psi_1}=\sqrt{q}\ket{\phi_1}+\sqrt{1-q}\ket{\phi_2},
\label{p1}
\end{equation}
and
\begin{equation}
p_{2}=\frac{1}{2},\,\,\, \ket{\Psi_2}=\sqrt{q}\ket{\phi_1}-\sqrt{1-q}\ket{\phi_2}.
\label{p2}
\end{equation}
\ensuremath{\hfill \Box}

\section{Numerical evidence for the validity of Theorem 2 for matrices with a rank larger than $2$ }

After presenting a proof for  Eq.~(\ref{e2b}) 
for the rank-$2$ case, in this section we examine numerical evidence for larger systems.
Thus, our approach is similar to the one followed in Ref.~\cite{HW97}, where the concurrence formula has been derived for rank-$2$ mixed states and for mixed states with a higher rank it has been supported by numerical evidence. 

We show a method to compute the infimum over convex decompositions in Eq.~(\ref{MAX}) by mapping this problem to an optimization over symmetric separable states  \cite{W89}. While the optimization for symmetric separable states can be computed only for $d=2,$ we get a lower bound on the infimum by optimizing for symmetric quantum states with a positive partial transpose (PPT), since states with a positive partial transpose are a superset of separable states \cite{PPT}. We will also consider a smaller superset, the set of quantum states with a PPT symmetric extension \cite{PPTsymext}. Note that the optimization for density matrices and operators of dimension $d$ is equivalent to optimization over density matrices of rank-$d$ allowing for density matrices and operators of any size.

The optimization problem for  $F_Q^{\max}[\varrho,A]$ defined in Eq.~(\ref{MAX}) can be rewritten as follows.
\begin{eqnarray}
F_Q^{\max}[\varrho,A]
&&=4\bigg(\exs{A^2}_{\varrho}-\sup_{\{p_{k},\ket{\Psi_{k}}\}}\sum_k p_{k} \exs{A}^2_{\Psi_k}\bigg),\nonumber\\
\end{eqnarray}
where $\{p_{k},\ket{\Psi_{k}}\}$ refers to a decomposition of $\varrho$ of the type Eq.~(\ref{decomp}).

Next, we will rewrite the term quadratic in expectation values as a term linear in expectation values of an operator acting on a bipartite system as
\begin{eqnarray}
&&F_Q^{\max}[\varrho,A]\nonumber\\
&&\;\;\;\;\;\;\;\;\;\;= 4\bigg(\exs{A^2}_{\varrho}-\sup_{\{p_{k},\ket{\Psi_{k}}\}}\sum_k p_{k} \exs{A\otimes A}_{\Psi_k\otimes \Psi_k}\bigg).\nonumber\\
\end{eqnarray}
Further transformations lead to a form in which instead of a sum of expectation values we have
an expectation value of a mixture as
\begin{eqnarray}
&&F_Q^{\max}[\varrho,A]\nonumber\\
&&\;\;\;\;\;\;\;\;\;\;= 4\bigg(\exs{A^2}_{\varrho}-\sup_{\{p_{k},\ket{\Psi_{k}}\}}
 \exs{A\otimes A}_{\sum_k p_k \vert \Psi_k \rangle \langle \Psi_k\vert ^{\otimes 2}}\bigg).\nonumber\\
 \end{eqnarray}
Hence, we arrive at a form based on an optimization over symmetric separable states 
 \begin{eqnarray}
F_Q^{\max}[\varrho,A]
&= &4\bigg(\exs{A^2}_{\varrho}-
\sup_{
   \mbox{
   \begin{tabular}{c}
      $\scriptstyle \varrho_{\rm ss} \in S_{\rm s},$\\
      $\scriptstyle {\rm Tr_1}(\varrho_{\rm ss})= \varrho$
   \end{tabular}
   }
}
\exs{A\otimes A}_{\varrho_{\rm ss}}\bigg)\nonumber\\
&=&2
\inf_{
   \mbox{
   \begin{tabular}{c}
      $\scriptstyle \varrho_{\rm ss} \in S_{\rm s},$\\
      $\scriptstyle {\rm Tr_1}(\varrho_{\rm ss})= \varrho$
   \end{tabular}
   }
}
\exs{(A\otimes 1-1\otimes A)^2}_{\varrho_{\rm ss}},\nonumber\\\label{Fbound}
\end{eqnarray}
where $S_{\rm s}$ is the set of symmetric separable states. States in $S_{\rm s}$ are mixtures of symmetric product states  \cite{W89,SymSep}
\begin{eqnarray}
\sum_k p_k \vert \Psi_k \rangle \langle \Psi_k\vert ^{\otimes 2}. \label{ssep}
\end{eqnarray}

\begin{table*}
\begin{tabular*}{\textwidth}{@{\extracolsep{\fill}\vspace{0.07cm}}llll}
\hline
\hline
\rule{0pt}{\normalbaselineskip}Task
& 
Largest Rel. Diff.
& 
Average Rel. Diff. 
& 
Standard Dev. of  Rel. Diff.\\
\hline
\rule{0pt}{\normalbaselineskip}\mbox{$2$SPPT, zeros in the diagonal} 
&$1.4793\times10^{-6}$ & $3.6194\times10^{-10}$& $2.2321\times10^{-8}$\\
$2$SPPT&$9.2505\times10^{-6}$ & $1.3313\times10^{-9}$& $9.8533\times10^{-8}$\\
$3$SPPT&$2.3552\times10^{-8}$ & $8.8486\times10^{-10}$& $9.9692\times10^{-10}$\\
$4$SPPT&$9.0345\times10^{-9}$ & $1.3448\times10^{-9}$& $9.8216\times10^{-10}$\\
$3$SE$_3$ &$7.7749\times10^{-8}$ & $4.2694\times10^{-10}$& $3.5161\times10^{-9}$\\
$4$SE$_3$ &$9.5506\times10^{-9}$ & $1.0429\times10^{-9}$& $1.0404\times10^{-9}$\\
\hline
\hline
\end{tabular*}
\caption{
 (Top line) Statistics of the relative difference between the quantum Fisher information and $\mathcal{B}_{\rm SPPT}(\varrho,A)$ for $10000$ random tests. For each test, a random $A$ with zero diagonal elements and a random $\varrho$ is generated. The largest relative difference, the average relative difference and the standard deviation  of the relative difference (i.e., the square root of the variance)
 is shown for two qudits for $d=2.$ SPPT indicates that the optimization is taken over PPT symmetric states.  (Second, third and fourth lines) The same for random $A$ with nonzero diagonal elements $d=2,3,4.$
(Bottom two lines) The same for $d=3,4$ for the relative difference between the quantum Fisher information and $\mathcal{B}_{{\rm SE}_3}(\varrho,A).$ SE$_3$ indicates that the optimization is taken over symmetric states with a PPT symmetric extension for $N=3$ qudits.}
\label{tab_numerics}
\end{table*}

It is not possible to compute the right-hand side of Eq.~(\ref{Fbound}) in general. However, 
it is possible to obtain a series of lower bounds on it based on the theory of PPT symmetric extensions \cite{PPTsymext}. Let us introduce the notation for such bounds 
\begin{eqnarray} \label{BEnnn}
 \mathcal{B}_{{\rm SE}_n}(A,\varrho) := 2
\inf_{
   \mbox{
   \begin{tabular}{c}
      $\scriptstyle\varrho_{{\rm SE}_{n}} \in S_{{\rm SE}_{n}}$\\
      $\scriptstyle{\rm Tr_1}(\varrho_{{\rm SE}_{n}})= \varrho $
   \end{tabular}
   }
}
\exs{(A\otimes 1-1\otimes A)^2}_{\varrho_{{\rm SE}_{n}}}, \nonumber\\\label{Bdef}
\end{eqnarray}
where $S_{{\rm SE}_n}$ are the symmetric bipartite states with a PPT symmetric extension of size $n.$ 
We note that $S_{{\rm SE}_2}$ is identical to the set of symmetric PPT states
\begin{equation}
S_{{\rm SE}_2} \equiv S_{\rm SPPT},
\end{equation}
hence we will use the two notations interchangeably in the rest of the paper. 
With the definition Eq.~(\ref{Bdef}), for $n>3$ we can write
\begin{eqnarray}
F_Q^{\max}[\varrho,A]&\ge&...\ge \mathcal{B}_{{\rm SE}_{(n+1)}}(A,\varrho) \ge \mathcal{B}_{{\rm SE}_{n}}(A,\varrho)  \nonumber\\
&\ge&...\ge\mathcal{B}_{{\rm SE}_{3}}(A,\varrho)\ge\mathcal{B}_{{\rm SPPT}}(A,\varrho).
\label{SE3boundB}
\end{eqnarray}
All these bounds can be obtained using semidefinite programming \cite{PPTsymext}.

The non-increasing series of bounds in Eq.~(\ref{SE3boundB}) exists since, in general,  states having a PPT symmetric extension for $n$ qudits are a subset of states having a PPT symmetric extension for $m<n$ qudits. 
In entanglement theory PPT states are typically separable and entangled states with a PPT property are 
very special \cite{PPT}. Thus, the bound $\mathcal{B}_{{\rm SPPT}}$ based on an optimization over PPT states is expected to be very close to the value obtained from optimization over separable states for most $A$'s and $\varrho$'s. The bound $\mathcal{B}_{{\rm SE}_{3}}$ computed 
based on an optimization over $S_{{\rm SE}_3}$ is expected to be even closer to the one obtained from an optimization over separable states.

We carried out numerical tests using random $A$ and $\varrho$ matrices. 
We considered $10000$ random trials for dimensions $d=2,3,4$ for computing the bound $\mathcal{B}_{\rm SPPT}$ defined in Eq.~\eqref{BEnnn}.  The matrix $A$ was generated as follows. First a matrix  $M$ was obtained such that all the real and imaginary parts of its elements were independent random numbers with a normal distribution centered around $0$ and with a variance $\sigma=1.$  Then, the Hermitian matrix $A$ was obtained from $M$ as $\frac{1}{2}(M+M^\dagger).$
The density matrix $\varrho$  was generated based on the method described in Ref.~\cite{RandomMat}. For each randomly chosen $A$ and $\varrho$ we determined the relative difference between $\mathcal{B}_{\rm SPPT}(\varrho,A)$ given in Eq.~(\ref{Bdef}) and $F_Q^{\rm BC}[\varrho,A].$ 
We used the program packages SeDuMi and YALMIP for semidefinite programming, while QUBIT4MATLAB V4.0 was used for calculations connected to quantum physics \cite{SeDuMi,YALMIP,QUBIT4MATLAB}.  
MATLAB uses double precision arithmetic.

To test our method, we considered first the $d=2$ case for $A$ matrices with zero diagonal elements. According to Theorem 2 and knowing that the set of PPT states and  the  set of separable states are the same for two qubits, $F_Q^{\rm BC}[\varrho,A]$ is equal to the bound  $\mathcal{B}_{\rm SPPT}(\varrho,A).$  The results are in the first row of Table~\ref{tab_numerics}. We found that the largest relative difference is below $2\times 10^{-6},$ while the average relative difference and the standard deviation of the relative difference are much smaller.
The nonzero value must be the result of the finite precision of the numerical calculations.

After testing the method for a case that have been proven analytically in Theorem 2, we also used it to
explore the cases that have not been proven yet.
We carried out calculations for $A$ matrices with nonzero diagonal elements for $d=2,3,4.$ The results of the numerical tests can be seen in the second, third and fourth rows of Table~\ref{tab_numerics}.
We also carried out similar numerical tests for dimensions $d=3,4$ for computing the bound 
$\mathcal{B}_{{\rm SE}_3}$based on an optimization for symmetric states with an $N=3$ PPT symmetric extension given in Eq.~\eqref{BEnnn}
and compare it to the quantum Fisher information. The results of the numerical tests can be seen in the bottom part of Table~\ref{tab_numerics}.

We can see that for all these trials, the relative difference between the two bounds and the value of $F_Q^{\rm BC}$ is smaller than $10^{-5},$ while the average relative difference is smaller than $10^{-8},$ and the standard deviation of the relative difference is smaller than $10^{-7}.$
Thus, the numerical tests suggest that it should be examined carefully through analytical calculations whether
\begin{eqnarray}
F_Q^{\max}[\varrho,A]=F_Q^{\rm BC}[\varrho,A] \label{FQmax}
\end{eqnarray}
is true in general.
Based on this strong evidence, we conjecture that this is the case.
 The semidefinite programs necessary to calculate $\mathcal{B}_{\rm SPPT}(\varrho,A)$
 and $\mathcal{B}_{{\rm SE}_3}(\varrho,A)$ are outlined in Appendix A.

\section{Maximal Quantum Fisher information within the set defined by Refs.~\cite{P03,GH09}}

In this section, first we will review how Refs.~\cite{P03,GH09}
define the variances and the types of generalized quantum Fisher information. We will denote
them by $\varrmh$ and $\hat{F}$, in order to distinguish them from
the generalized variances and various types of Fisher information defined in this paper.
Then, we will show that these definitions, apart from a constant factor,
fulfill Definition 1 and Definition 3. This is a surprise since these
quantities in Refs.~\cite{P03,GH09}
were defined based on ideas very different from the ones
presented in this paper. We also show that the set of variances and 
the types of quantum Fisher information satisfying Definition 1 and Definition 3
are a broader set than the corresponding quantities presented in Refs.~\cite{P03,GH09}.
Finally, we will show that the usual quantum
Fisher information given in Eq.~(\ref{eq:qF}), after appropriate normalization,
 is the largest even within the various types of quantum Fisher information considered in 
Refs.~\cite{P03,GH09}.

\subsection{Summary of the basic results of Refs.~\cite{P03,GH09}}

The basic idea of Refs.~\cite{P03,GH09} is that for each standard
matrix monotone function $f:\mathbb{R}^{+}\rightarrow\mathbb{R}^{+},$
a generalized variance and a corresponding quantum Fisher information
are defined.  The notion standard means that $f(1)=1$ and $f(t)=tf(t^{-1}).$
First, let us consider the generalized variances.

\textbf{Definition 5.}The variance according to Refs.~\cite{P03,GH09}
is defined as

\begin{equation}
\varrmh_{\varrho}^{f}(A)=\langle A
,\mathbb{J}_{\varrho}^{f}(A
)\rangle-({\rm Tr}\varrho A)^2
,\label{eq:Covq2}
\end{equation}
where $A$ is Hermitian, the scalar product is defined as $\langle A,B\rangle = {\rm Tr}(AB),$ and
\begin{equation}
\mathbb{J}^f_{\varrho}(A)=
f(\mathbb{\mathbb{L_{\varrho}}R_{\varrho}}^{-1}
)\mathbb{R_{\varrho}}
.
\end{equation}
where 
\begin{align*}
\mathbb{L_{\varrho}}(A) & =\varrho A,\quad\mathbb{R_{\varrho}}(A)=A\varrho.
\end{align*}
 
Computing $\mathbb{J}_{\varrho}(A)$ can be simplified knowing that for $f(x)=\sum_k c_k x^k$ we have \cite{P03,GH09}
\begin{equation}
\mathbb{J}^f_{\varrho}(A)=\sum_k c_k \varrho^k A \varrho^{(1-k)}. \label{rar}
\end{equation}
Thus, Eq.~(\ref{rar}) can be used to calculate $  \mathbb{J}_{\varrho}(A)$  for any polynomial of $x.$
 
Moreover,  it is also useful to define the mean based on $f$ as
\begin{equation}
m_{f}(a,b)=af\left(\frac{b}{a}\right)
\end{equation}
 and use it instead of $f.$  The $f(1)=1$ normalization condition corresponds to  the condition 
$m_{f}(a,a)=a$ for the means. The $f(t)=tf(t^{-1})$ requirement corresponds to $m_{f}(a,b)=m_{f}(b,a).$
A list of quantum Fisher information generated by various  well-known means $m_{f}(a,b)$ can be found in Refs.~\cite{P03,GH09,DG11}.

After we discussed the generalized variances, we will turn our attention
to the quantum Fisher information.

\textbf{Definition 6.} The quantum Fisher information according
to Refs.~\cite{P03,GH09} is defined as
\begin{equation}
\hat{F}^{f}(\varrho;A)={\rm Tr}(A\mathbb{J}_{f}^{-1}(\varrho)A).\label{eq:qFdef}
\end{equation}
Again, the most important types of generalized quantum Fisher information
correspond to famous means appearing often in mathematics.

For the arithmetic mean $m_{f}(a,b)=\frac{a+b}{2},$ the quantum
Fisher information is defined as
\begin{equation}
\hat{F}^{{\rm BC}}(\varrho;A)=\int_{0}^{\infty}{\rm Tr}[\exp(-t\varrho/2)A\exp(-t\varrho/2)A]dt.
\end{equation}
Other means generate other types of quantum Fisher information. Details can be found in Refs.~\cite{P03,GH09,DG11}.

\subsection{The usual quantum Fisher information is the maximal one within the set of quantum Fisher information defined in Refs.~\cite{P03,GH09}}

After reviewing the results of Refs.~\cite{P03,GH09}, we will connect the results of the present paper
to that of Refs.~\cite{P03,GH09}. In particular, we will show that the usual quantum Fisher information, Eq.~(\ref{eq:qF}),
is the largest within the set of quantum Fisher informations defined by Refs.~\cite{P03,GH09}, if they are normalized such that for pure states they equal four times the usual variance given in Eq.~(\ref{eq:usualvar}). 

In order to proceed, we will now prove the following lemma concerning the behavior of the family of variances defined in Refs.~\cite{P03,GH09} for pure states.

\textbf{Lemma 3} \emph{ For pure states we have
\begin{equation}
\varrmh_{\varrho}^{f}(A)=2 m_{f}(1,0)\times\va A.\label{eq:mf10}
\end{equation}
Thus, for pure states the variance of Refs.~\cite{P03,GH09} equals
the usual variance times a constant. } 

\emph{Proof.} It can be shown that Eq.~\eqref{eq:Covq2} can be rewritten
as \cite{GH09} 
\begin{eqnarray}
\varrmh_{\varrho}^{f}(A)&=&\sum_{ij}m_{f}(\lambda_{i},\lambda_{j})\vert A_{ij} \vert^2-\vert\sum\lambda_{i}A_{ii}^{\text{}}\vert^{2},
\label{eq:qCov}
\end{eqnarray}
where $\lambda_{j}$ are the eigenvalues of $\varrho$, and $A$ is
given in the basis of the eigenvectors of the density matrix. Here we took advantage of the fact that 
$m_{f}(a,a)=1.$
One can see that for pure states ($\lambda_{1}=1,$ $\lambda_{k}=0$
for $k\ge2$) we have
\begin{equation} \label{f2}
\varrmh_{\varrho}^{f}(A)=2\sum_{j>1}m_{f}(1,0)\vert A_{1j} \vert^2.
\end{equation}
Hence, Eq.~\eqref{eq:mf10} follows. Note that there is a factor of $2$ in Eq.~\eqref{f2} since
$m_{f}(\lambda_{i},\lambda_{j})$ in Eq.~\eqref{eq:qCov} is nonzero for the pure state considered for $i=1, j>1$ and for $ j=1, i>1.$
\ensuremath{\hfill \Box}

Next, we will turn our attention to the quantum Fisher information. We define the following notation  
\begin{equation}
\hat{F}{}_{Q}^{f}[\varrho,A]=\hat{F}^{f}(\varrho;i[\varrho,A]).\label{eq:irhoA}
\end{equation}
With this definition, for the arithmetic mean $m_{f}(a,b)=\frac{a+b}{2},$
$\hat{F}{}_{Q}^{f}[\varrho,A]$ equals the usual quantum Fisher information given in
Eq.~\eqref{eq:qF} denoted in the physics literature as $F_{Q}[\varrho,A]$
\cite{PS01}. 

It is instructive to analyze the differences between the quantum information $\hat{F}^{f}(\varrho;A)$ appearing in Refs.~\cite{P03,GH09} and, in general,
 in the mathematics literature and $F_{Q}^f[\varrho,A],$ which appears in the physics literature.
The difference between the two notions is that 
$1/\hat{F}^{f}(\varrho;A)$  bounds the squared uncertainty when the small parameter $t$ of 
\begin{equation}
\varrho_{\rm output}^{\rm Math.}(t)=\varrho+At
\end{equation}
 is estimated \cite{P03}. Here $A$ is a matrix, and hence in this case the output density matrix is a linear function of $t.$  In contrast,
 $1/F_{Q}^f[\varrho,A]$ bounds the squared uncertainty when
estimating $\theta$ in 
\begin{equation}\label{rhot}
\varrho_{\rm output}(t)=\exp(-iA\theta)\varrho\exp(+iA\theta),
\end{equation}
as mentioned in the introduction. The definition of $\hat{F}^{f}(\varrho,A)$ is not clear for 
pure states, as for such states $\varrho+At$ is aphysical for $t>0$ or $t<0,$
while  Eq.~(\ref{rhot}) describes a physical dynamics that leads to 
a valid density matrix for all $t.$

Next, we will prove a connection between the generalized quantum Fisher information and the usual variance for pure states.

\textbf{Lemma 4.} \emph{For pure states, the generalized quantum Fisher information $\hat{F}_{Q}^{f}[\varrho,A]$ is proportional to the usual variance given in Eq.~(\ref{eq:usualvar})
\begin{equation}\label{FFQQ}
\hat{F}_{Q}^{f}[\varrho,A]=\frac{2\va A}{m_{f}(1,0)}.
\end{equation}}

\emph{Proof.} Equation~\eqref{eq:qFdef} leads to \cite{GH09}
\begin{equation}\label{Ff}
\hat{F}^{f}(\varrho;A)=\sum_{i,j}\frac{1}{m_{f}(\lambda_{i},\lambda_{j})}\vert A_{ij}\vert^{2},
\end{equation}
where $\lambda_{j}$ are the eigenvalues of $\varrho$, and $A$ is
given in the basis of the eigenvectors of the density matrix. Substituting Eq.~\eqref{Ff} into
Eq.~\eqref{eq:irhoA}, we obtain
\begin{equation}
\hat{F}_{Q}^{f}[\varrho,A]=\sum_{i,j}\frac{(\lambda_{i}-\lambda_{j})^{2}}{m_{f}(\lambda_{i},\lambda_{j})}\vert A_{ij}\vert^{2}.\label{eq:Fq_m}
\end{equation}
For $m_{f}(a,b)=\frac{a+b}{2},$ we obtain the usual quantum Fisher information given in
Eq.~\eqref{eq:qF}. The lemma can be proven substituting $\lambda_{1}=1,$
$\lambda_{k}=0$ for $k\ge 2$ into Eq.~\eqref{eq:Fq_m}. \ensuremath{\hfill \Box}

Next, we will present the central theorems of this section.

\textbf{Theorem 4.} \emph{After appropriate normalization, the generalized
variances defined in Refs.~\cite{P03,GH09} fulfill Definition 1.
However, the opposite is not true. Not all generalized variances that
fulfill Definition 1 belong to the generalized variances defined in
Refs.~\cite{P03,GH09}.}

\emph{Proof.} It has been proven that the generalized variances
of Refs.~\cite{P03,GH09} are concave \cite{P10}. Based on this
fact and Lemma 3, we see that 
\begin{equation}\label{nvar}
{\rm var}_{\varrho}^{f}(A)=\frac{\varrmh_{\varrho}^{f}(A)}{2m_{f}(1,0)}
\end{equation}
is a generalized variance in the sense of Definition 1. At this point
the question arises: Are the the two definitions equivalent? We find
that this is not the case. Eq.~\eqref{eq:example2} is a generalized
variance according to Definition 1, but cannot be written in the form
Eq.~\eqref{eq:Covq2} remembering that $\mathbb{J}_{\varrho}(A)$ is defined in Eq.~(\ref{rar}). This is essentially due to the fact that $\varrho^2$ appears in the definition Eq.~\eqref{eq:example2}.
On the other hand, Refs.~\cite{P03,GH09}
defines a family of variances based
on a single-variable function $f,$ while Definition 1 does not provide
an explicit formula for obtaining such quantities. \ensuremath{\hfill \Box}

\textbf{Theorem 5. }\emph{After appropriate normalization, the types of quantum
Fisher information defined in Refs.~\cite{P03,GH09} fulfill Definition
3.}

\emph{Proof.} It has been proven that the types of generalized quantum Fisher information
of Refs.~\cite{P03,GH09} are convex \cite{P10}. 
Based on Lemma 4,
we can also see that

\begin{equation}\label{XXX}
{F}_{Q}^{f}[\varrho,A]=2m_{f}(1,0)\hat{F}_{Q}^{f}[\varrho,A]
\end{equation}
is a generalized quantum Fisher information in the sense of Definition
3. \ensuremath{\hfill \Box}

Next, we discuss, which quantum Fisher information is the largest from the family considered in Refs.~\cite{P03,GH09}.
It has been proven that $\hat{F}^{f}(\varrho;A)$ is the smallest for $f(x)=\frac{1+x}{2},$ that is for the usual quantum Fisher information \cite{P03,GH09}. Moreover, it has been also found that the quantum Fisher information for $f(x)=\frac{2x}{1+x},$ i.e., for the harmonic mean, is the smallest  \cite{P03,GH09}.
This can be seen as follows. In the expression for the quantum Fisher information, Eq.~(\ref{Ff}),  appears the $m_f(a,b)$ mean. The arithmetic mean $m_f(a,b)=\frac{a+b}{2}$ is the largest among the means, while the harmonic mean is the smallest.

With a similar argument, we can prove that $\hat{F}^{f}[\varrho,A]$ defined in Eq.~(\ref{eq:Fq_m}) is the largest for  $f(x)=\frac{1+x}{2},$ i.e., for the  arithmetic mean. It can also be proven that it is the smallest for the harmonic mean.

According to Theorem 4, the variance fitting Definition 1 is of the form Eq.~(\ref{nvar}).
Moreover, according to Theorem 5, the types of quantum Fisher information fitting the Definition 3 are of the form Eq.~(\ref{XXX}). Both differ from the definitions of Refs.~\cite{P03,GH09} by a normalization factor.
For which $f(x)$ is the quantum Fisher information ${F}_{Q}^{f}[\varrho,A]$
the largest, if it is normalized such that for pure states it equals
$4(\Delta A)^2$? In order to answer this question, based on the derivation above, let us write down explicitly the variances and types of quantum Fisher information defined by  Refs.~\cite{P03,GH09}, together with the appropriate normalization factors such that for pure states they equal to the usual variance, Eq.~(\ref{eq:usualvar}), and to four times the usual variance, respectively. Based on Eqs.~\eqref{eq:mf10}, \eqref{eq:qCov}, \eqref{FFQQ} and \eqref{eq:Fq_m},
the two definitions are as follows.
\begin{eqnarray}
{\rm var}_{\varrho}^{f}(A)&=&\frac{1}{2}\sum_{i, j}\frac{m_{f}(\lambda_{i},\lambda_{j})}{m_{f}(1,0)}\vert A_{ij}\vert^{2}-
\vert\sum\lambda_{i}A_{ii}^{\text{}}\vert^{2},\nonumber\\
{F}_{Q}^{f}[\varrho,A]&=&2\sum_{i,j}\frac{m_{f}(1,0)}{m_{f}(\lambda_{i},\lambda_{j})}{(\lambda_{i}-\lambda_{j})^{2}}\vert A_{ij}\vert^{2}.\label{eq:Fq_m2}\nonumber\\\label{vF}
\end{eqnarray}
For each mean $m_f(a,b),$  Equation \eqref{vF}  provides a generalized variance fulfilling Definition 1 and a 
generalized quantum Fisher information fulfilling Definition 3.
It can be seen that the same $m_f(a,b)$ corresponds to the minimal variance and to the maximal quantum Fisher information. However, from Theorem 1, we already know that the minimal variance is the usual variance given in 
Eq.~(\ref{eq:usualvar}), corresponding to  $f(x)=\frac{1+x}{2},$ i.e., to the arithmetic mean $m_f(a,b)=\frac{a+b}{2}.$ Hence, Theorem 3 follows.

\section{Conclusions}

We have defined generalized variances and different types of quantum Fisher information.
We have shown that the smallest variance is the usual variance since it is its own
concave roof. We have also shown that the largest quantum Fisher information
is the usual quantum Fisher information for rank-$2$ density matrices
and for operators that have zero diagonal elements in the eigenbasis of the density matrix.
In other words, the quantum Fisher information is 4 times the convex roof
of the variance for such matrices. We found strong numerical evidence that the quantum Fisher
information is very close to the convex roof even
for matrices with non-zero diagonal elements and for matrices with a larger rank.
We related our findings to the generalized variances
and types of quantum Fisher information defined in Refs.~\cite{P03,DG11}. 
We showed that the 
usual quantum Fisher information is the largest within the various types of
generalized quantum Fisher information of Refs.~\cite{P03,GH09},
if appropriate normalization is applied.
Hence we conjectured that the usual quantum Fisher information is four times the convex roof of the variance for
density matrices and Hermitian operators of any size.

In the future, it would be important to prove Conjecture 1.
Since convex roofs appear in 
entanglement theory very often, our findings help to
establish further connections between the quantum Fisher information and entanglement theory \cite{PS01,FisherMultipartieEntanglement,FisherMultipartieEntanglement2,FisherMultipartieEntanglement3,HG10,FD11,FD12,FD12B}.
For instance, it would be important to examine the consequences of our findings
concerning the detection of entanglement
with variances \cite{EntUnc} or the quantum Fisher information \cite{PS01,FisherMultipartieEntanglement,FisherMultipartieEntanglement2,FisherMultipartieEntanglement3}, 
or the estimation of the quantum Fisher information
based on measurements \cite{private,unpub}. 
It would also be important to point out connections
to the recent findings in Refs.~\cite{EM11,ED12},
which, in another context, also relate the quantum Fisher information to convex roof constructions.

\section{Acknowledgments}

We thank O.~G\"uhne, P.~Hyllus, 
M.~Kleinmann, M.~W.~Mitchell, and Z.~Zimbor\'as for discussions. 
We thank K. Demkowicz-Dobrza\'nski for drawing our attention to Ref.~\cite{EM11}.
We acknowledge the
support of the ERC Starting Grant GEDENTQOPT, CHIST-ERA QUASAR,
the Spanish MICINN (project No. FIS2009-12773-C02-02 and No. FIS2012-36673-C03-03),
the Basque Government (project No. IT4720-10), and the National Research Fund of Hungary
OTKA (Contract No. K83858 and No. K104206).

\appendix

\section{Semidefinite programs used for obtaining the global optimum}

Here we show how to compute $\mathcal{B}_{{\rm SPPT}}(\varrho,A)\equiv\mathcal{B}_{{\rm SE}_2}(\varrho,A)$ and $\mathcal{B}_{{\rm SE}_3}(\varrho,A)$ defined in Eq.~\eqref{BEnnn} using semidefinite programming.

Semi-definite programming can be used to look for the infimum over symmetric PPT states, i.e, finding $\mathcal{B}_{{\rm SE}_2}(\varrho,A)$ given in Eq.~\eqref{BEnnn}. In this way we obtain an upper bound on the infimum for separable states as has also been used in Ref.~\cite{NJP09}. The corresponding task can be formulated as a
standard semidefinite
program as
\begin{equation}
  \begin{array}{ll}
    \text{{\bf minimize}} &2\exs{(A\otimes 1-1\otimes A)^2}_{\varrho_0}, \\
    \\
    \text{{\bf subject to}} & \varrho=\sum_{m,n} \vert s_m (2,d) \rangle \langle s_n (2,d) \vert (\varrho_s)_{mn},\\
                             & \varrho_0\ge 0, \;\; \trace(\varrho_0)=1, \\
                             &  \varrho_0^{T_1}\ge 0, \;\;   \trace_2(\varrho_0)= \varrho. \\
  \end{array}
\end{equation}
Here $\vert s_n (2,d)\rangle$ denotes the basis states of the symmetric subspace for $2$ qudits of dimension $d.$ In order to ensure that the optimization is only over symmetric states, we define the density matrix $\varrho_s,$
which is a two-qudit symmetric state given in the basis of the symmetric states. $\varrho_0$ is a density matrix given in the product basis, and $T_1$ means partial transpose according to the first subsystem.

Semi-definite programming can also be used to look for the infimum over symmetric states that have a PPT symmetric extension over $N$ qudits appearing in $\mathcal{B}_{{\rm SE}_N}(\varrho,A)$ defined in Eq.~\eqref{BEnnn}. 
The corresponding task can be formulated as a semidefinite
program as
\begin{equation}
  \begin{array}{ll}
    \text{{\bf minimize}} & 2\exs{(A\otimes 1-1\otimes A)^2}_{\varrho_{12}}, \\
    \\
    \text{{\bf subject to}} & \varrho_0=\sum_{m,n}  \vert s_m(N,d) \rangle \langle s_n(N,d) \vert (\varrho_s)_{mn},\\
                             & \varrho_0\ge 0, \;\; \trace(\varrho_0)=1,\\
                            & \varrho_0^{T_{1,2,..,M}}\ge 0 \text{ for } M=1,2,..., \lfloor\frac{N}{2}\rfloor,\\
                                    & \varrho_{12}= \trace_{3,4,..,N} (\varrho_0),  \;\;  \trace_2(\varrho_{12})= \varrho. \\
  \end{array}
\end{equation}
where  $\varrho_{12}$ is the two-qudit reduced state of  $\varrho_0$ and   $\lfloor x\rfloor $ is the integer part of $x.$
Now  $\varrho_{\rm s}$ and $\varrho_0$ are $N$-qudit states.
Here $\vert s_n(N,d) \rangle$ denotes the basis states of the symmetric subspace for $N$ qudits of dimension $d.$ 
For $N$ qubits, these are 
\begin{equation}
\vert s_n(N,2) \rangle=
\binom{n}{N}^{-\frac{1}{2}}\sum_{k}\mathcal{P}_{k}(\vert1
\rangle^{\otimes n}\vert0\rangle^{\otimes (N-n)}),
\end{equation}
for $n=0,1,...,N,$ where the summation is over all different permutations.
Now $T_I$ means partial transposition according to the group of qudits given in $I.$
Such an optimization has already been used in  Ref.~\cite{TG09} for studying the entanglement properties
of symmetric quantum states.

\end{document}